\documentclass{article}

\usepackage[preprint]{neurips_2023}
\usepackage{graphicx}
\usepackage{amsmath} 
\usepackage{multirow}
\usepackage{xcolor}
\usepackage{hyperref}
\hypersetup{
    colorlinks=true,
    urlcolor=blue!60!black
}

\usepackage[utf8]{inputenc} 
\usepackage[T1]{fontenc}    
\usepackage{hyperref}       
\usepackage{url}            
\usepackage{booktabs}       
\usepackage{amsfonts}       
\usepackage{nicefrac}       
\usepackage{microtype}      
\usepackage{xcolor}         
\usepackage{enumitem}
\usepackage{parskip}  
\title{\textbf{Mini-Omni2}: Towards Open-source GPT-4o with Vision, Speech and Duplex Capabilities}

\author{%
  Zhifei Xie$^\spadesuit$$^\clubsuit$ \\
  \texttt{xzf24@mails.tsinghua.edu.cn} \\
  \and
\textbf{Changqiao Wu$^\spadesuit$}\thanks{Corresponding author. Work done during Zhifei Xie's internship at Inspirai.}\\
\texttt{wuchangqiao@inspirai.com} \\
\and
  $^\spadesuit$ Inspirai \quad
 $^\clubsuit$ Tsinghua University \\
 \href{https://github.com/gpt-omni/mini-omni2}{\textcolor{blue!60!black}{\texttt{https://github.com/gpt-omni/mini-omni2}}}
  }
\begin{document}
\maketitle

\begin{abstract}

GPT-4o, an all-encompassing model, represents a milestone in the development of large multi-modal language models. It can understand visual, auditory, and textual modalities, directly output audio, and support flexible duplex interaction. Models from the open-source community often achieve some functionalities of GPT-4o, such as visual understanding and voice chat. Nevertheless, training a unified model that incorporates all modalities is challenging due to the complexities of multi-modal data, intricate model architectures, and training processes. In this paper, we introduce \textbf{Mini-Omni2}, a visual-audio assistant capable of providing real-time, end-to-end voice responses to visoin and audio queries. By integrating pretrained visual and auditory encoders, \textbf{Mini-Omni2} maintains performance in individual modalities. We propose a three-stage training process to align modalities, allowing the language model to handle multi-modal inputs and outputs after training on a limited dataset. For interaction, we introduce a command-based interruption mechanism, enabling more flexible interaction with users. To the best of our knowledge, \textbf{Mini-Omni2} is one of the closest reproductions of GPT-4o, which have similar form of functionality, and we hope it can offer valuable insights for subsequent research.

\end{abstract}

\begin{figure}[h]
\begin{center}
    \includegraphics[width=0.8\textwidth]{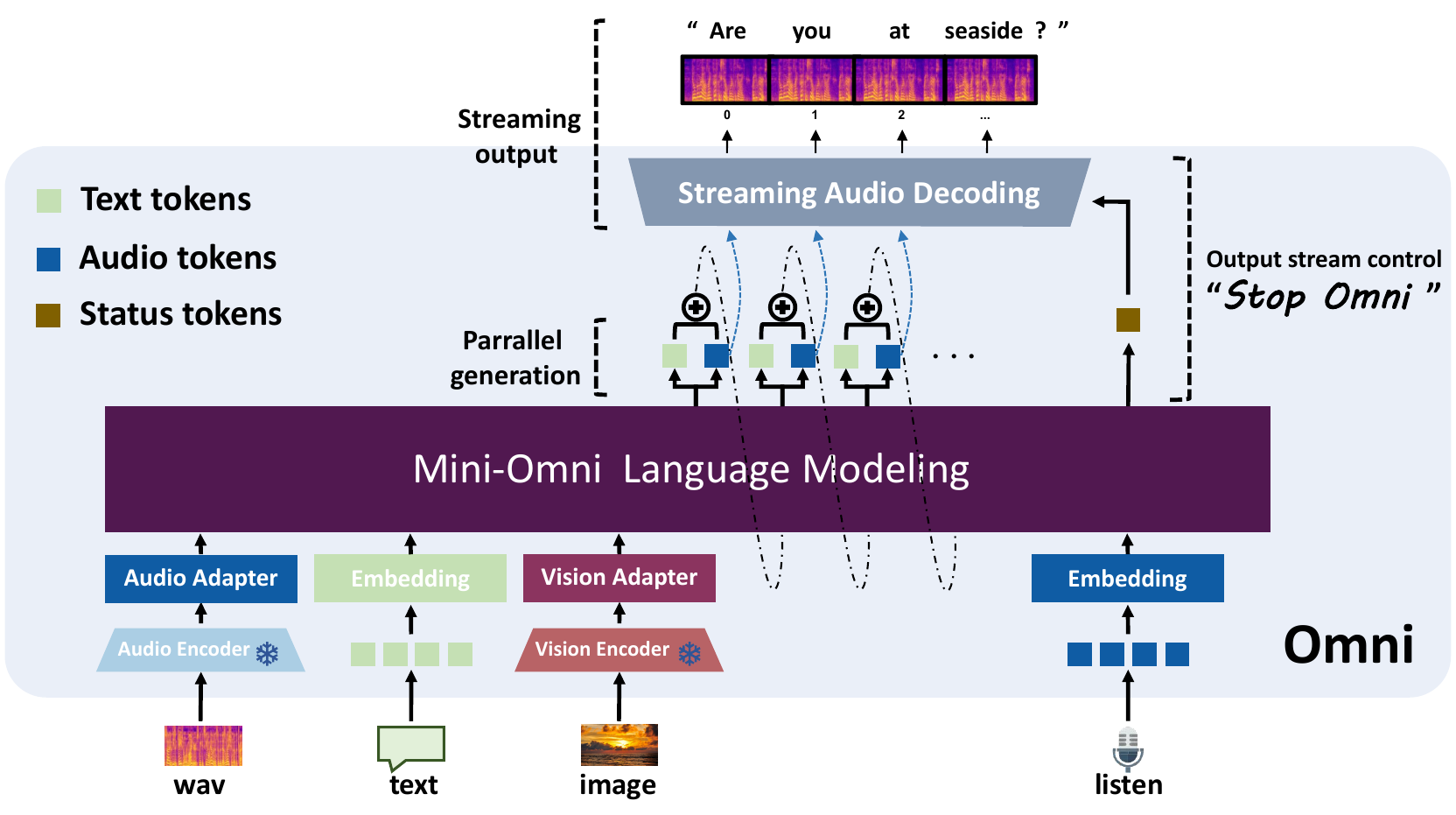}
\end{center}
\caption{The \textbf{Mini-Omni2} model architecture.}
\end{figure}

\section{Introduction}
GPT-4o\citep{gpt4o} represents a milestone in the development of multi-modal large language models, particularly evident in three aspects: \textbf{(1)} its powerful capabilities in multi-modal question answering; \textbf{(2)} its ability to transcend traditional text-based input and output, enabling the understanding and generation of multi-modal content; and \textbf{(3)} its flexible interaction mode with interruption mechanisms, which facilitates a more natural and fluid human-computer interaction. However, the GPT-4o model is not open-sourced to the public, and its technical specifications remain undisclosed. To date, mainstream methods predominantly involve employing various pre-trained encoders to obtain textual outputs for specific modalities, such as visual and audio inputs, and utilizing cascading techniques to integrate a text-to-speech (TTS) module that replicates GPT-4o's speech output capabilities, thereby simulating its multi-modal functionalities. Achieving end-to-end multi-modal understanding and output remains a challenging task.

Recently, as the capabilities of language models such as Llama3.2\citep{llama3.1} continue to expand, researchers have begun to explore multi-modal approaches to achieve the performance of GPT-4o. However, these research outcomes often focus only on specific functionalities of GPT-4o, such as vision-text understanding (LLava\citep{llava}, Flamingo\citep{flamingo}), audio comprehension (Qwen2-audio\citep{qwen2audio}), multi-modal understanding (VITA\citep{vita}), and speech-to-speech dialogue (Mini-Omni\citep{mini-omni}, Llama-Omni\citep{llama-omni}, Moshi\citep{moshi}). However, integrating text, vision, and speech modalities remain challenging.

In our view, the current challenges in achieving interaction across three modalities involve the following aspects: \textbf{(1) Model capability} — GPT-4o requires a unified model that comprehensively understands all modalities while maintaining robust performance across wide range of tasks; \textbf{(2) direct inference output capabilities in multi-modal contexts} — our recent work Mini-Omni\citep{mini-omni} has addressed how to enhance the model's streaming output abilities in audio, laying the groundwork for \textbf{Mini-Omni2}'s voice interaction capabilities; \textbf{(3) substantial data requirements} — training for GPT-4o necessitates the integration of data across visual, audio, and textual modalities, with quantities increasing exponentially compared to previous efforts; \textbf{(4) the design of flexible interaction methods} — GPT-4o's full-duplex capability is also a notable feature.

In this paper, we introduce \textbf{Mini-Omni2} as a continuation of Mini-Omni, employing a single model to end-to-end simulate the visual, speech, and textual capabilities of GPT-4o, enhanced by a unique command-based interruption mechanism. Consistent with Mini-Omni, we retain Qwen2\citep{qwen2} as the foundational model, leveraging this compact architecture to achieve comprehensive multi-modal understanding and real-time streaming speech inference across the three modalities. Furthermore, we enable the model to receive external audio inputs in real time, simulating its "auditory" perception and controlling the speech output stream based on content semantics.  The model architecture of \textbf{Mini-Omni2} is illustrated in Figure 1. As an end-to-end model, we enhance data utilization efficiency and demonstrate the generalizability of the \textbf{Mini-Omni2} algorithm by directly employing the classic pre-trained visual encoder CLIP\citep{clip} and the encoder component of the speech recognition model Whisper\citep{whisper} as feature extractors for visual and audio inputs. The features from the pre-trained encoders and the text embedding are concatenated to form the model's input. Due to challenges related to understanding capabilities, we did not adopt a token-in-token-out paradigm. Moreover, utilizing a delayed parallel output approach for text and audio, the model can response instantly with audio like GPT-4o.

In \textbf{Mini-Omni2}, we propose an efficient training approach based on a limited amount of data, aiming to enable the model's training methods to assist other multi-modal models in modality expansion. Thus, we avoided blindly expanding the dataset exponentially and instead sought to develop a multi-modal extension method using minimal new data. We employed a three-phase training process for modality expansion, alignment, and joint training. Initially, the \textbf{Mini-Omni2} model underwent adapter training using speech recognition and image caption datasets, thereby broadening the scope of multi-modal understanding. Next, \textbf{Mini-Omni2} was trained for text output in question-answering tasks across modalities, allowing the adapter-based output features to align with text embedding for effective question answering. In the third phase, we focused on multi-modal output capability by incorporating audio output and training for auditory capabilities like interruption.
\begin{figure}[htbp]
\begin{center}
    \includegraphics[width=1.0\textwidth]{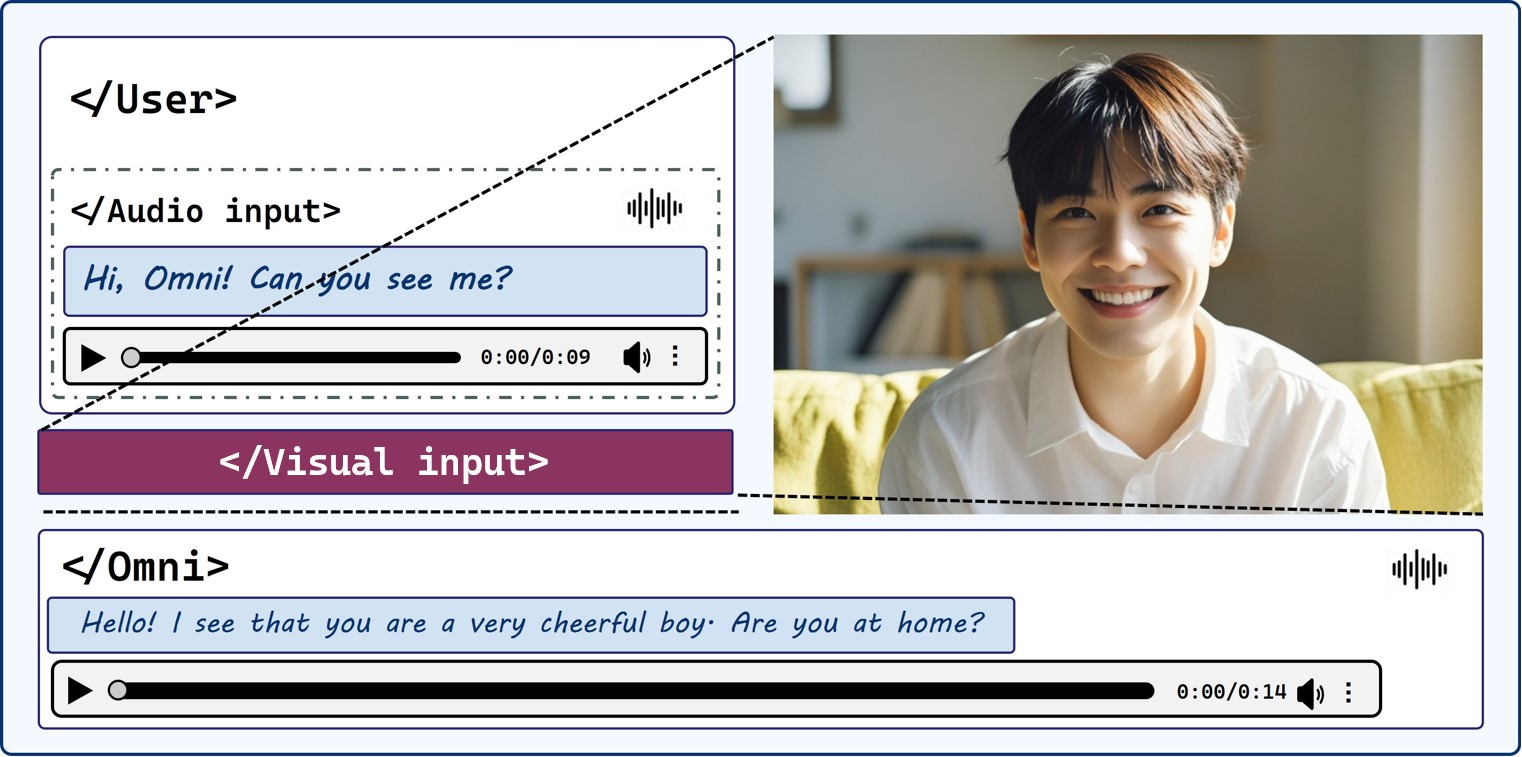}
\end{center}
\caption{\textbf{Mini-Omni2} now supports streaming speech responses for image, audio and text inputs. 
}
\end{figure}

With respect to the model's capabilities in voice interaction, \textbf{Mini-Omni2} continues to utilize the \textbf{SNAC} tokenizer\citep{snac} to ensure high-quality speech output. However, based on our observations, we believe that the current full-duplex training is still not sufficiently stable. 
Therefore, we contend that interruptions based on input semantic information are essential for achieving stable and flexible human-machine interaction. We enable the model to perform real-time encoding of its received "auditory" waveforms using SNAC, generating tokens that allow it to control its own output during each generation. As a demonstration, we construct data using the phrase "stop omni," employing frame-level \textbf{irq} and \textbf{n-irq} special tokens to control the generating process. 

To evaluate the multi-modal interaction capabilities of \textbf{Mini-Omni2}, we first empirically tested its performance on traditional visual and auditory tasks, verifying that the model maintains consistency with the original model in basic tasks such as image caption and speech recognition. Next, we conducted a series of additional experiments to test the model's response speed and perform some case studies.


\noindent \textbf{In summary, we make the following contributions:}
\begin{itemize}[leftmargin=*]
    \item We introduce \textbf{Mini-Omni2}, the first open-source multi-modal language model with capabilities in vision, speech, text and an auditory interruption mechanism. To the best of our knowledge, it is one of the most similar end-to-end models to the GPT-4o's functionalities. Figure 2 shows the demo of the model as a visual voice assistant.
    \item
    We propose a novel training pipeline based on the modal expansion method from the previous Mini-Omni. This pipeline encompasses three training phases, allowing the text model to first align responses to multi-modal inputs, and ultimately extend outputs to the speech modality in the final phase, employing a delayed parallel generation algorithm for real-time speech output.
    \item We explored a command-based interruption method, utilizing streaming tokens as input and constructing training data to enable the model to control its audio output stream based on external semantic cues. And all the synthetic data will be open-sourced.
\end{itemize}

\section{Related Work}

\textbf{Large Vision Language Models} Recent vision-language models are developing rapidly and were among the first modalities to combine with large language models. The foundational work began with CLIP\citep{clip}, which is also used as vision encoder in our work. Subsequent works typically employ a vision encoder, an adapter as an intermediate layer, and a large language model as the architecture to enable the LLM to understand and reason about visual inputs. Classic works include BLIP\citep{blip}, BLIP2\citep{blip}, Llava\citep{llava}, Qwen-VL\citep{qwenvl}, Qwen2-VL\citep{qwen2vl}, InstructBLIP\citep{instructblip}, MiniGPT-4\citep{minigpt4}, GPT-4V\citep{gpt4v} from OpenAI, Gemini\citep{gemini} from Google, and Llama-3.2\citep{llama3.1} from Meta. Researchers are also exploring other directions, such as higher-resolution vision encoders like InternLM-XComposer2-4KHD\citep{internlm} and using MOE architectures, as in works like CogVLM\citep{cogvlm}. The method used in this paper is the most classical, which is similar to Llava\citep{llava}.

\textbf{Audio Language Modeling} With the further development of large multi-modal models, speech signals have also been discretized into tokens, enabling understanding and reasoning in a manner similar to text models. Important works include speech synthesis models like VALL-E\citep{valle}, music generation models like MusicGen\citep{MusicGEN}, as well as voice interaction works like AudioPaLM\citep{audiopalm} and LauraGPT\citep{lauragpt}. Just recently, researchers have explored methods for speech-to-speech interaction, with works such as Mini-Omni\citep{mini-omni}, Llama-Omni\citep{llama-omni}, and Moshi\citep{moshi}. Speech tokenization is also an important direction for generating stable and information-rich tokens, with recent works like Speechtokenizer\citep{speechtokenizer}, Google USM\citep{googleusm}, and EnCodec\citep{encodec}.

\textbf{Multi-modal Interaction Model} With the emergence of GPT-4o, researchers have begun working on end-to-end multi-modal models for voice chat. Early works include Spectron\citep{Spectron} and SpeechGPT\citep{speechgpt}, which use the A-T-T-A method to achieve speech-in and speech-out in an end-to-end manner. Mini-Omni\citep{mini-omni} introduced a method for parallel generation of text and audio, enabling the model to directly start reasoning in audio. Both Moshi\citep{moshi} and Llama-Omni\citep{llama-omni} used similar approaches. LSLM\citep{LSLM} and Moshi explored the full duplex interaction capability by combining the speaking and listening signals as input. VITA\citep{vita} can understand all modalities but only outputs text. The AnyGPT\citep{anygpt} project aims to achieve full multi-modal understanding and generation. This work is a continuation of Mini-Omni, aiming to realize multi-modal input and low-latency parallel speech-text output with duplex capability.

\section{\textbf{Mini-Omni2}}

The model architecture of \textbf{Mini-Omni2} is illustrated in Figure 1. In addition to the text embedding module, \textbf{Mini-Omni2} employs the visual component of CLIP and Whisper-small as encoders for visual and auditory modalities, resulting in highly efficient data utilization during training and minimizing extensive pre-training efforts. 
Additionally, \textbf{Mini-Omni2} features real-time duplex capability, providing greater flexibility in model interactions. 
This section includes 3.1, which discusses the model architecture; 3.2, which presents the modeling methods for input and output streams; and sections 3.3 and 3.4, which detail training methods and interuption manner, respectively.

\begin{figure}[htbp]
\begin{center}
    \includegraphics[width=1.0\textwidth]{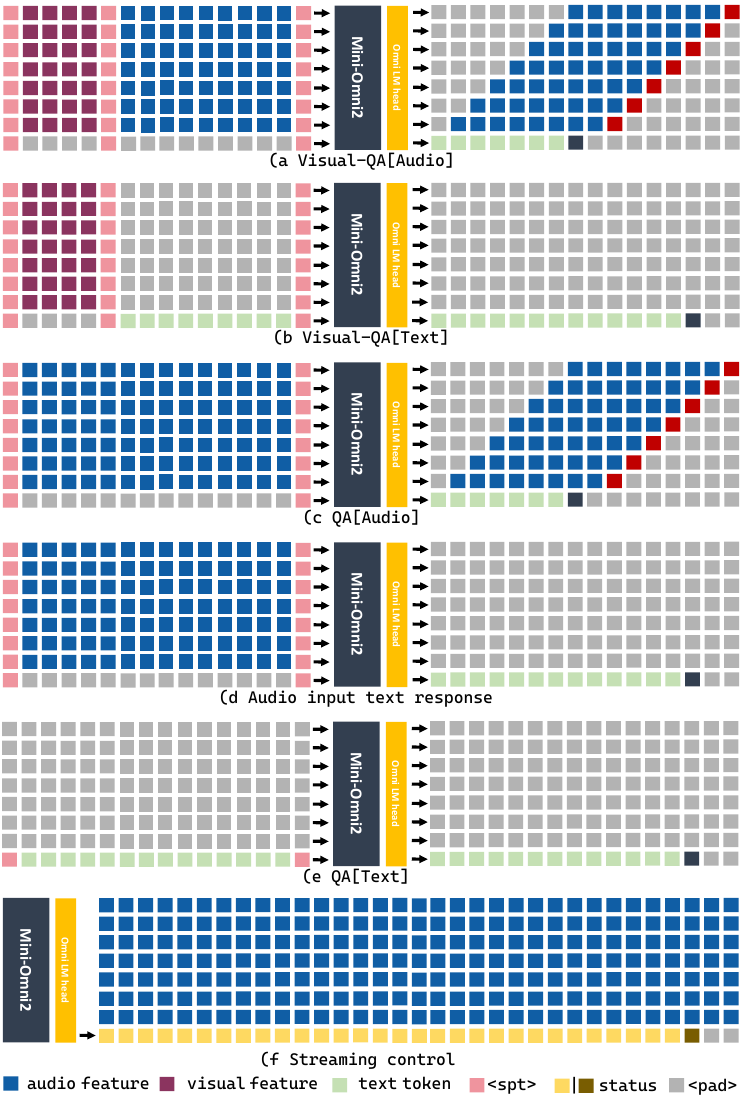}
\end{center}
\caption{Schematic diagram of multi-layer tokens for input and output of the main task model of \textbf{Mini-Omni2}.
}
\end{figure}

\subsection{Architecture}

\textbf{Visual Encoder} - We utilize the visual component of CLIP, specifically the ViT-B/32 model, as the visual encoder, which converts incoming images to a feature sequence of length 49 for the image patches and a global semantic feature. \textbf{Mini-Omni2} concatenates these to form a raw feature sequence of length 50, employing a single-layer LlamaMLP\citep{llama} as the vision adapter. 

\textbf{Audio Encoder} - In the encoder section, we continue our previous work by using the Whisper-small model as the audio encoder. We opted not to adopt a token-in-token-out modeling approach for audio input and output for two reasons. \textbf{(i) Strong semantic alignment in speech recognition}. The Whisper model, proposed by OpenAI, is trained on thousands of hours of datasets, demonstrating exceptional robustness. Furthermore, we unexpectedly found that Mini-Omni exhibits an understanding of Chinese data, despite not being trained on any Chinese datasets. We believe this is due to the Whisper model's capability to automatically align audio from different languages, tones, and noise levels that convey the same meaning, thereby enabling the model to focus on the user's intention. 
\textbf{(ii) Unstable open-source audio tokens}. We observed a phenomenon where a) the audio loss of \textbf{Mini-Omni2} remains high during training, and b) the tokens for a segment of audio can vary significantly based on the content at both ends. We argue that tokens are insufficient for reliably conveying the content of speech input, as evidenced by the poor performance of ASR comparing to semantic features like Whisper.

\textbf{Language Model} - \textbf{Mini-Omni2} uses the Qwen2-0.5B base version as its foundational language model. We have ported the Llama-based Qwen2 model using the LitGPT\citep{litgpt-2023} training framework, employing the configuration of the 0.5B model as the base language model. For the parallel generation of the multi-layer codebook shown in Figure 3, we expanded the vocabulary of the Qwen2 model by adding 7 × 4160 sub-LM-heads, as illustrated in Figure 4, resulting in a vocabulary size of 181,120.

\begin{figure}[h]
\begin{center}
    \includegraphics[width=1.0\textwidth]{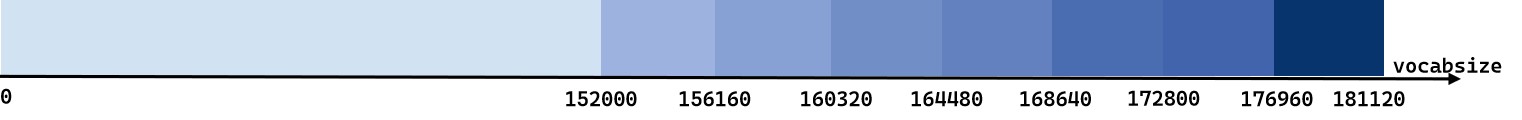}
\end{center}
\caption{The special multi-layer vocabulary construction of \textbf{Mini-Omni2}.
}
\end{figure}

\subsection{Multimodal Languague Modeling}

\textbf{Multimodal Modeling} - Consider \( Y = (y_i \in \mathcal{V}_{\text{txt}} \mid i = 1, \ldots, t_{\text{txt}}) \) as a text utterance from a vocabulary \(\mathcal{V}_{\text{txt}}\) with length \( t_{\text{txt}} \). The probability of \( Y \) can be expressed as \( p(Y) = \prod_{i=1}^{t_{\text{txt}}} p(y_i \mid y_1, \ldots, y_{i-1}) \). Now, when dealing with a continuous speech signal, we can convert it into discrete speech tokens (\(\text{dst}\)), represented as \( D = (d_i \in \mathcal{V}_{\text{dst}} | i = 1, \cdots , t_{\text{dst}}) \) using a audio tokenizer. In this context \(\mathcal{V}_{\text{dst}}\) is the vocabulary of discrete speech tokens. These discrete speech tokens can be treated as spoken language within \(\mathcal{V}_{\text{dst}}\) and modeled in a manner similar to text. We combine text and speech in a new vocabulary \(\mathcal{V}_{\text{voxt}}\) by \( \mathcal{V}_{\text{voxt}} = \mathcal{V}_{\text{txt}} \cup \mathcal{V}_{\text{dst}} \). Additionally, we introduce visual features \(V \in \mathcal{F}_{\text{vis}}\), 
where \(\mathcal{F}_{\text{vis}}\) represents the continuous features extracted from the image. Therefore, we can model the probability of both speech, text, 
where \( Z = (z_i \in \mathcal{V} | i = 1, \cdots , t) \). This probability is expressed as \( p(Z) = \prod_{i=1}^{t} p(z_i \mid z_1, \cdots, z_{i-1}, V)\), where \( Z \) represents discrete speech tokens \( D(\mathcal{V} = \mathcal{V}_{\text{dst}}) \), text tokens \( Y(\mathcal{V} = \mathcal{V}_{\text{txt}}) \), and continuous video features \( V(\mathcal{F} = \mathcal{F}_{\text{vis}}) \), or various combinations of \( Y \), \( D \), and \( V \). For the audio and text tokens generated simultaneously, the negative log-likelihood loss can be formulated as in Equation (1).

\begin{equation}
\mathcal{L}(T,A,V|C) = \sum_{j=1}^{m}\sum_{i=1}^{n_{j}} \log P(T_{i,j}, A_{i,j} | T_{<i,j}, A_{<i,j}, V_j; X_j)
\end{equation}

where \( T \), \( A \) are the text-audio output pairs in the training corpus \( C \), and \( m \) is the number of training examples. \( X_j \) and \(V_j\) is the input condition of the \( j \)-th example, \( n_j \) is the max number of tokens of sample \( T_j \) and \( A_j \), and \( T_{i,j} \) and \( A_{i,j} \) represent the \( i \)-th text token and audio token of the \( j \)-th sample.

\textbf{Multi-modal token-Mixed Input} - The modeling of input and output tokens for some of the model's main tasks is illustrated in Figure 3. In this section, we will discuss the model's inputs and outputs. Since the model incorporates multiple LM-heads, it generates multiple sequences in an auto-regressive manner. As a result, the model also takes multiple sequences as inputs. The input sequences can include a mixed input from a minimum of one modality to a maximum of three modalities. In this subsection, we will discuss the methods for modality mixing.

\begin{itemize}[leftmargin=*]
    \item \textbf{\textit{Visual-[Audio|Text] Input}} Our experiments indicate that the Transformer architecture is easier to train and generates more natural responses when auto-regressive tasks are connected with semantic information. Therefore, as shown in Figure 3 (a), we first place the visual features processed by the vision adapter, followed by the Whisper features processed by the audio adapter. Finally, at the position where a response needs to be generated auto-regressively, we place a special token for the response. The total length is approximately 50(CLIP feature length) + \( L_a \)(Whisper feature length).

    \item \textbf{\textit{Single Modality Input}} Single-modal inputs may consist of visual, speech, or text inputs. We place the features of both visual and audio modalities across layers 1 to 7. These features will be replicated to enhance their prominence when averaged across all layer features. Notably, when only a single modality's features are input without the control of a special token, the default tasks are image caption, speech-to-text question answering, and text-to-text question answering.
\end{itemize}

\begin{figure}[htbp]
\begin{center}
    \includegraphics[width=1.0\textwidth]{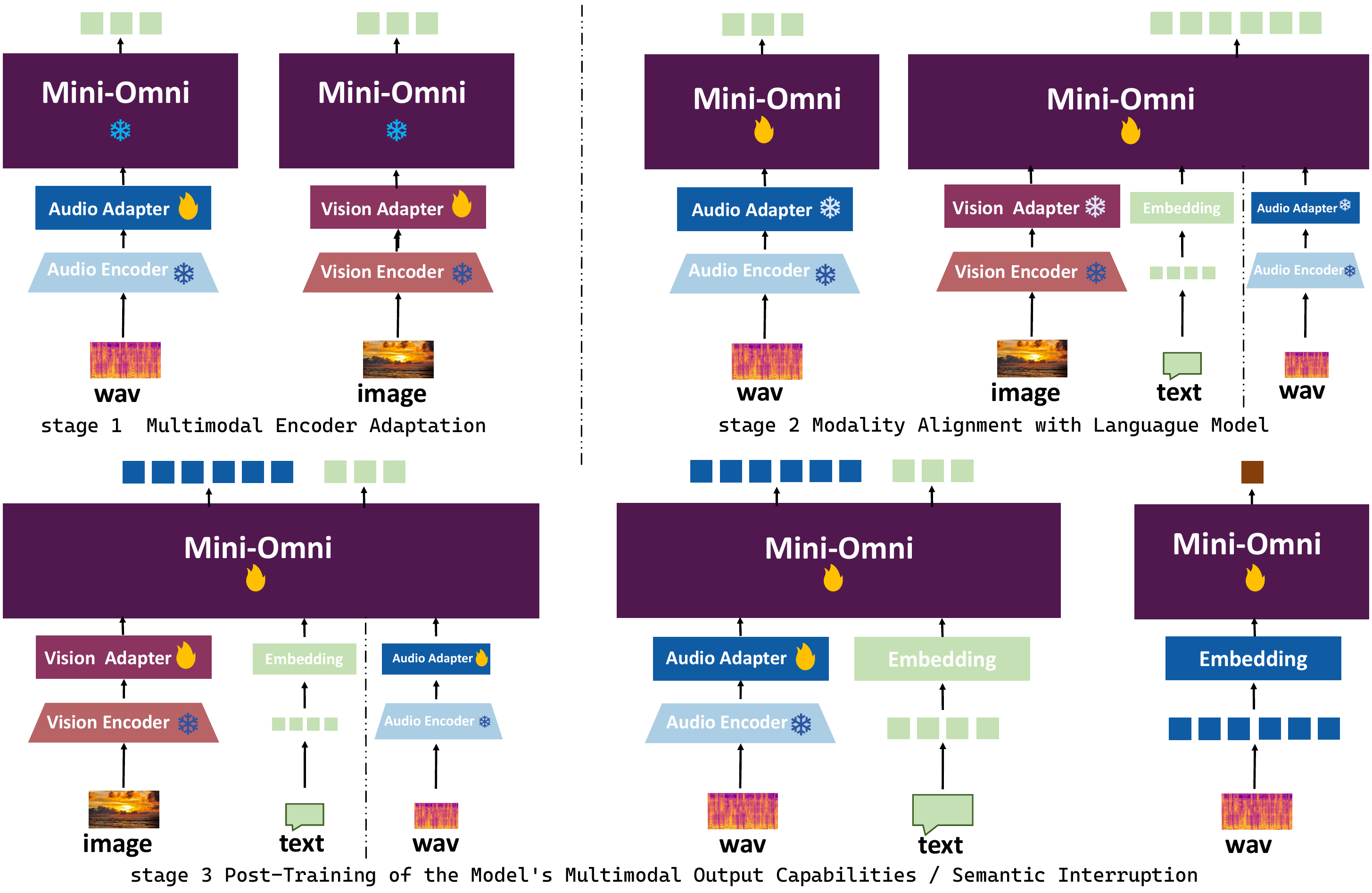}
\end{center}
\caption{\textbf{Mini-Omni2}'s three-stage training phases}
\end{figure}

\textbf{Text-Audio Parrallel Decoding} In \textbf{Mini-Omni2}, we essentially retain the output strategy of Mini-Omni, employing the Text-Instruct Delay Parallel Decoding algorithm to enhance audio generation. This approach utilizes text-audio parallel decoding to simultaneously generate audio and text tokens, leveraging text-to-speech synthesis for real-time output.
We continue the parallel generation method introduced by MusicGen\citep{MusicGEN}, utilizing SNAC as the audio encoder, which comprises seven complementary token layers. In a single step, we generate eight tokens, including text, while maintaining a one-step delay between layers. Furthermore, we incorporate a Batch approach that involves two samples: one requiring both text and audio responses and the other necessitating a text-only response. By discarding the text token from the first sample and embedding the output from the second sample into the first, we effectively transfer the model’s text-based capabilities to audio tasks, significantly enhancing reasoning abilities with minimal resource overhead. We have provided detailed explanations of the specific technical details in the Mini-Omni\citep{mini-omni}.

Overall, we have introduced our modeling approach for three-modal inputs and two-modal outputs within a single model. Through these methods, the model can accomplish eight reasonable multi-modal tasks, with some of the primary tasks illustrated in Figure 3, showcasing all the multilayer tokens generated during a single inference.

\subsection{Training Strategies}
In this section, we will introduce the training phase of the \textbf{Mini-Omni2} model. The overall training process of \textbf{Mini-Omni2} is illustrated in Figure 5. The training process is divided into three stages, with multitask training employed in each stage. In the figure, except for Stage 1, a foundational text-to-text task is additionally incorporated but not explicitly depicted. We categorize the entire training process into three stages:

\begin{itemize}[leftmargin=*]
    \item \textbf{\textit{Multimodal Encoder Adaptation}}   In the first stage, we employ a rapid, small-scale training focused solely on the weights of the linear layer connecting the language model and the encoder. The objective of Stage 1 is to ensure that the multi-modal features received by the model closely resemble the characteristics of text tokens as represented in the model's embedding layer. We believe this approach offers two primary advantages: 1. It allows the model to concentrate on logical reasoning in modality-specific question answering during subsequent training. 2. It minimizes the parameter changes in the language model's core that would otherwise result from adapting to other modalities.

    \item \textbf{\textit{Modality Alignment}}   In stage 2, the primary task of the model training is to transfer the question-answering ability based on text input to question-answering abilities based on images and audio. In this step, the adapters trained in stage 1 are temporarily frozen, and the weights of the language model are involved in the training. At this stage, all tasks do not involve audio responses. For tasks like image-based and audio-based QA, only text-based responses are generated to establish the model's foundational logical capabilities. The speech output is simply an extension of this logical ability into different modalities.

    \item \textbf{\textit{Post training}}   In Stage 3, the task of the model is to extend the output modality to include audio response generation. As shown in Figure 5, the model will be trained on all tasks from Stage 1 and Stage 2, with audio token outputs for all question-answering tasks. Additionally, the model will learn interruption mechanism, an algorithm introduced in the next section.
\end{itemize}

\subsection{Duplex Interaction}
A real-time conversation model needs to have duplex capability in order to enable more flexible interactions. However, this interruption mechanism should not be a simple VAD (Voice Activity Detection)-based one, but rather a system that can determine whether the user intends to interrupt the model. Additionally, the model's ability should be highly robust, capable of handling various external situations (e.g., noise, other conversations, and unrelated sounds). We explore this functionality with command-based task, where the model will stop talking immediately when user speaks "Stop Omni". Furthermore, this approach can be naturally extended to incorporate more sophisticated semantic interruption mechanisms through the development of more contextually appropriate interruption datasets.

\textbf{Background Noise Selection:} (1) We randomly utilized a variety of speech recognition samples from the Libri-tts dataset as the original human noise data samples. (2) We employed samples from the MUSAN\citep{snyder2015musan} dataset, which includes music, human voices, white noise, and urban noise.

\textbf{Semantic Interruption Construction:} We synthesized "Stop Omni" phrases with random voice timbres, which were subsequently mixed with noise. The specific data construction methods are introduced in the next section.

Combining the aforementioned data, the model will receive long sequences of data containing "Stop Omni" phrases amidst various noises. The model will generate two types of state tokens in real time: {irq} and {n-irq}, representing the intention of the user to interrupt and not to interrupt, respectively. During inference, when the model output {irq} token, it will stop the generating process and start to listen the new question. For this task, we use tokens as input to enhance the model's real-time processing capabilities. 

\section{Data and Evaluation}
In this section, we introduce the data used for training \textbf{\textbf{Mini-Omni2}} and present some initial evaluation results. We will provide a more detailed explanation of the composition and construction process of the data for each modality. In the experimental results section, we only showcase a few application cases and basic capability assessments. More comprehensive experiments related text and vision tasks will be updated shortly.

\subsection{Datasets}
The training data for the \textbf{Mini-Omni2} model is primarily sourced from five components, as shown in Table 1. (1) Textual Question-Answering Data: Throughout all training stages, whenever the language model weights were unfrozen for training, textual question-answering data was included to maintain the model's reasoning ability. We used the first 1.5 million question-answer pairs from the Open-Orca dataset. (2) Speech Recognition Data: Speech recognition data was used to continuously maintain the model’s semantic understanding of external spoken input. We primarily utilized the LibriTTS, VCTK, and Multilingual LibriSpeech datasets. (3) Spoken Question-Answering Data: We did not use a standalone spoken dataset; instead, synthetic data was employed for training. The spoken question-answering data was derived from the Moss-002-sft dataset. (4) Image Question-Answering Data: We used 400,000 samples (caption and instruction) from the ALLaVA-4V dataset. (5) Voice Assistant Data: To make the model’s responses more aligned with the style of voice assistants, we continuously used the VoiceAssistant-400K dataset introduced in Mini-Omni.

\begin{table}[h]
\centering
\begin{tabular}{lllll}
\toprule
\textbf{Task} & \textbf{Stages}   & \textbf{Dataset} & \textbf{Modality} & \textbf{items} \\ 
\midrule
          & & Libritts \citep{libritts}                   & A1|T1       & 586 h \\
ASR & 1,2,3 & VCTK \citep{vctk}                           & A1|T1       & 44 h  \\
          & & Multilingual LibriSpeech \citep{mls}        & A1|T1       & 8000h \\ \midrule
          
 Text QA & 2,3& Open-Orca \citep{openorca}                & T1|T2       & 2000K  \\ \midrule
 
 Audio QA & 2,3 & Moss-002-sft-data \citep{moss}         & A1|T1|A2|T2 & 1500K \\ \midrule
Visual QA & 2,3 & ALLaVA-4V \citep{moss}                 & V|A1|T1|A2|T2 & 800K \\ \midrule
 
          & & Alpaca-GPT4 \citep{Alpacagpt4}              & A1|T1|A2|T2 & 55k \\
          & & Identity finetune \citep{identity_finetune} & A1|T1|A2|T2 & 2k \\
          & & QAassistant \citep{qa-assistant-2}          & A1|T1|A2|T2 & 27k \\
  voice QA  & final & Rlhf \citep{rlhf}                   & A1|T1|A2|T2 & 367k \\ 
          & & Trivia-singlechoice \citep{t-sim}           & A1|T1|A2|T2 & 17k \\
          & & Trivia-Multichoice \citep{t-mul}            &A1|T1|A2|T2  & 20k \\
          & &  OpenAssistant \citep{open-assis}           & A1|T1|A2|T2 & 2k \\
\bottomrule
\end{tabular}
\caption{The datasets and their usage for training \textbf{\textbf{Mini-Omni2}.}}
\end{table}

\subsection{Training Parameters} 
The \textbf{Mini-Omni2} model completed all training steps on eight A100 GPUs. During the adapter training stage, learning rates ranged from 2e-5 to 1e-3, while training the language model used learning rates between 2e-6 and 2e-4. The final fine-tuning was conducted with learning rates ranging from 2e-6 to 2e-5. A cosine scheduler was employed, with 1,500 warm-up steps and a global batch size of 192. Each stage was trained for one epoch using the full dataset. The scale of the vision and audio encoders was described earlier, and the language model used was the Qwen2-0.5B base model.
All model adapters used Llama-MLP with an intermediate size of 4,864.

\subsection{Data Construction } 
\textbf{Spoken Dialogue Data}: We used our speech recognition dataset as a random voice timbre library. To ensure robust training, a random sample from this dataset was selected as a voice prompt for the input of all spoken dialogue data, and CosyVoice\citep{cosyvoice} was employed for zero-shot speech synthesis. For the output of all question-answering data, the same voice timbre was used from an internal TTS system.

\textbf{Interruption Data}: First, the noise data is stream-encoded and decoded to simulate real-time streaming input to the model. Then, a random segment of the noise data is extracted. At the end of this segment, a "Stop Omni" phrase is inserted, generated with a random voice timbre in the same manner as the dialogue data. Finally, an additional "tail" of 0-10 seconds is appended to the end of this segment. In terms of labeling, all data before the tail is labeled as "n-irq", while the tail segment is labeled as "irq", indicating that the model should be interrupted.



\subsection{Experimental Results} 
Currently, we provide the accuracy of \textbf{Mini-Omni2} in speech recognition to evaluate the model's speech understanding ability, and we present some practical cases. For model experience and more cases, please follow our github repositories.

\begin{table}[h]
\centering
\begin{tabular}{lcccc}
\toprule
\textbf{Method} & \textbf{test-clean} & \textbf{test-other} & \textbf{dev-clean}  & \textbf{dev-other} \\ 
\midrule
Wav2vec2-base \citep{wav2vec}  &  6.0 &13.4 & - & - \\
VITA \citep{vita} &  8.14 & 18.41 & 7.57 & 16.57  \\
Whisper-small* & 4.4  & 10.1 &  4.6  & 10.3 \\
Mini-Omni &  4.5 & 9.7 & 4.6 &  9.2  \\
\textbf{Mini-Omni2} &  4.8 & 9.8 & 4.7 &  9.4  \\
\bottomrule
\end{tabular}
\caption{Comparison of the model's ASR with the base model used. (* our reproduced evaluation result.)}
\end{table}


According to the speech recognition results in Table 2, it can be observed that the accuracy of \textbf{Mini-Omni2} shows a slight decline after adding the visual modality compared to Mini-Omni. This phenomenon may be attributed to the relative reduction in the proportion of data. Moreover, in comparison with the decoder of the whisper module employed by the model, the \textbf{Mini-Omni2} model outperforms Whisper on the librispeech-other dataset. This demonstrates that our training process has enhanced the robustness of the model in speech recognition.

\subsection{Case Study}
Here we present some use cases from \textbf{Mini-Omni2}.
\begin{figure}[h]
\begin{center}
    \includegraphics[width=1.0\textwidth]{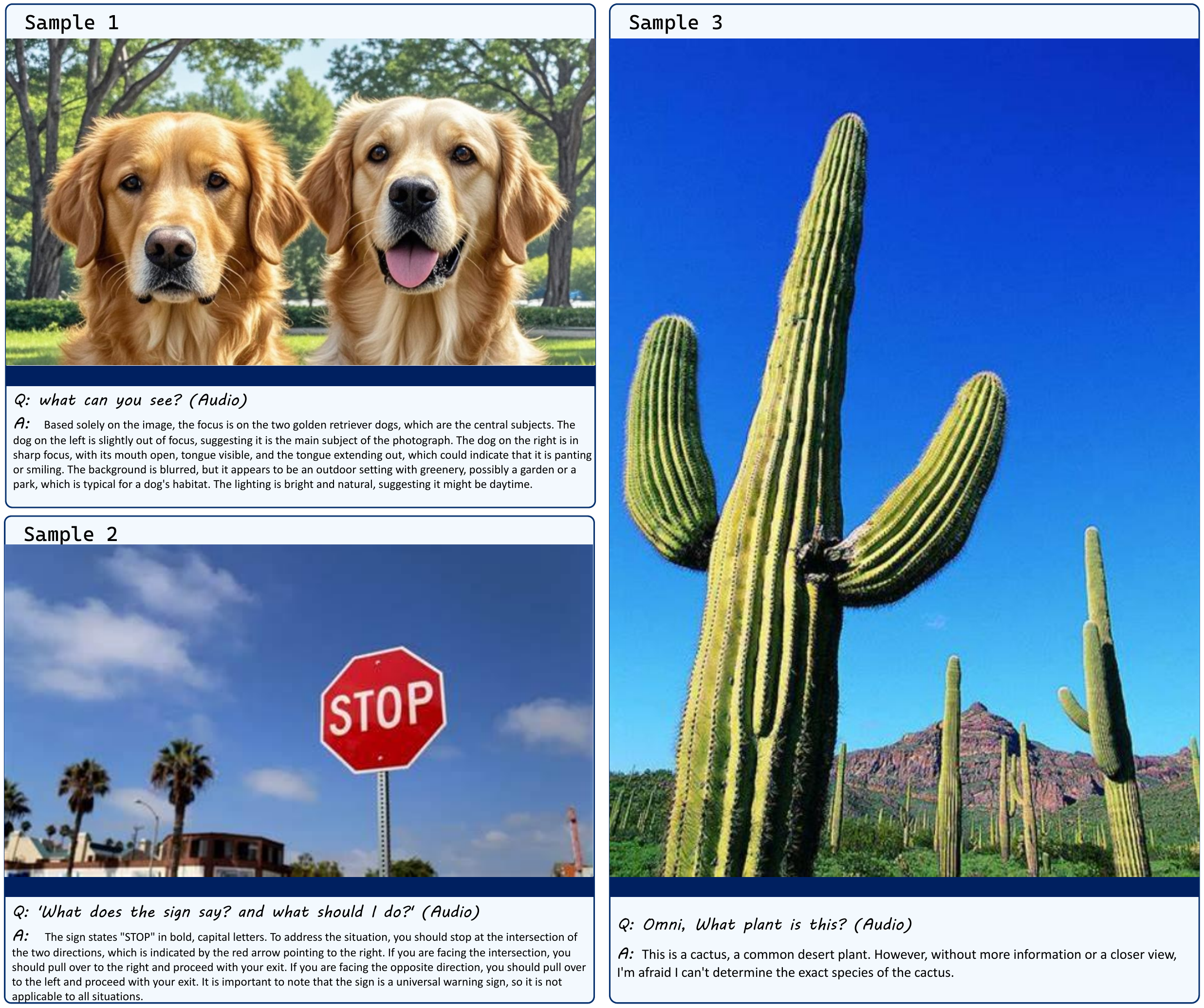}
\end{center}
\caption{Use cases of \textbf{Mini-Omni2}}
\end{figure}

\section{Limitations}
We believe the following aspects are worth exploring and improving:
\textbf{1.} Scaling of model and data size. \textbf{Mini-Omni2} aims to train small models with limited resources, and we believe that more data and compute can greatly enhance its capabilities.
\textbf{2.} Improve style control and diversity of audio output (emotion, naturalness, timbre, accent, and singing).
\textbf{3.} A richer mechanism for semantic interruptions.

\section{Conclusion}
In this paper, we present \textbf{Mini-Omni2}, a unified multi-modal language model with capabilities in text, speech, vision, end-to-end streaming audio output, and duplex interaction. Our goal is to reproduce an open-source GPT-4o model, and to our best knowledge, our work is also one of the closest in terms of functionality. We use multiple pretrained encoders as the vision and speech encoders and align them with the language model to extend the modalities. Furthermore, we propose a three-phase modality alignment and expansion training process to achieve the desired capabilities of the model.
We also explore a robust method for duplex interaction modeling and introduce our data construction and interruption mechanism. 
All models and datasets will be open-sourced, and we hope \textbf{Mini-Omni2} can serve as a reference for future research.

\bibliographystyle{plainnat} 
\bibliography{ref}

\begin{thebibliography}{50}
\providecommand{\natexlab}[1]{#1}
\providecommand{\url}[1]{\texttt{#1}}
\expandafter\ifx\csname urlstyle\endcsname\relax
  \providecommand{\doi}[1]{doi: #1}\else
  \providecommand{\doi}{doi: \begingroup \urlstyle{rm}\Url}\fi

\bibitem[AI(2023)]{litgpt-2023}
Lightning AI.
\newblock Litgpt.
\newblock \url{https://github.com/Lightning-AI/litgpt}, 2023.

\bibitem[Alayrac et~al.(2022)Alayrac, Donahue, Luc, Miech, Barr, Hasson, Lenc, Mensch, Millican, Reynolds, et~al.]{flamingo}
Jean-Baptiste Alayrac, Jeff Donahue, Pauline Luc, Antoine Miech, Iain Barr, Yana Hasson, Karel Lenc, Arthur Mensch, Katherine Millican, Malcolm Reynolds, et~al.
\newblock Flamingo: a visual language model for few-shot learning.
\newblock \emph{Advances in neural information processing systems}, 35:\penalty0 23716--23736, 2022.

\bibitem[Anthropic(2024)]{rlhf}
Anthropic.
\newblock https://huggingface.co/datasets/anthropic/hh-rlhf, 2024.

\bibitem[Baevski et~al.(2020)Baevski, Zhou, Mohamed, and Auli]{wav2vec}
Alexei Baevski, Yuhao Zhou, Abdelrahman Mohamed, and Michael Auli.
\newblock wav2vec 2.0: A framework for self-supervised learning of speech representations.
\newblock \emph{Advances in neural information processing systems}, 33:\penalty0 12449--12460, 2020.

\bibitem[Bai et~al.(2023)Bai, Bai, Yang, Wang, Tan, Wang, Lin, Zhou, and Zhou]{qwenvl}
Jinze Bai, Shuai Bai, Shusheng Yang, Shijie Wang, Sinan Tan, Peng Wang, Junyang Lin, Chang Zhou, and Jingren Zhou.
\newblock Qwen-vl: A frontier large vision-language model with versatile abilities.
\newblock \emph{arXiv preprint arXiv:2308.12966}, 2023.

\bibitem[Chen et~al.(2023)Chen, Chu, Gao, Li, Hu, Zhou, Xu, Ma, Wang, Zheng, et~al.]{lauragpt}
Qian Chen, Yunfei Chu, Zhifu Gao, Zerui Li, Kai Hu, Xiaohuan Zhou, Jin Xu, Ziyang Ma, Wen Wang, Siqi Zheng, et~al.
\newblock Lauragpt: Listen, attend, understand, and regenerate audio with gpt.
\newblock \emph{arXiv preprint arXiv:2310.04673}, 2023.

\bibitem[Chu et~al.(2024)Chu, Xu, Yang, Wei, Wei, Guo, Leng, Lv, He, Lin, et~al.]{qwen2audio}
Yunfei Chu, Jin Xu, Qian Yang, Haojie Wei, Xipin Wei, Zhifang Guo, Yichong Leng, Yuanjun Lv, Jinzheng He, Junyang Lin, et~al.
\newblock Qwen2-audio technical report.
\newblock \emph{arXiv preprint arXiv:2407.10759}, 2024.

\bibitem[Copet et~al.(2024)Copet, Kreuk, Gat, Remez, Kant, Synnaeve, Adi, and D{\'e}fossez]{MusicGEN}
Jade Copet, Felix Kreuk, Itai Gat, Tal Remez, David Kant, Gabriel Synnaeve, Yossi Adi, and Alexandre D{\'e}fossez.
\newblock Simple and controllable music generation.
\newblock \emph{Advances in Neural Information Processing Systems}, 36, 2024.

\bibitem[Dai et~al.(2023)Dai, Li, Li, Tiong, Zhao, Wang, Li, Fung, and Hoi]{instructblip}
Wenliang Dai, Junnan Li, Dongxu Li, Anthony Meng~Huat Tiong, Junqi Zhao, Weisheng Wang, Boyang Li, Pascale Fung, and Steven Hoi.
\newblock Instructblip: Towards general-purpose vision-language models with instruction tuning, 2023.
\newblock URL \url{https://arxiv.org/abs/2305.06500}.

\bibitem[datashare(2024)]{vctk}
datashare.
\newblock https://datashare.ed.ac.uk/handle/10283/2651, 2024.

\bibitem[D{\'e}fossez et~al.(2022)D{\'e}fossez, Copet, Synnaeve, and Adi]{encodec}
Alexandre D{\'e}fossez, Jade Copet, Gabriel Synnaeve, and Yossi Adi.
\newblock High fidelity neural audio compression.
\newblock \emph{arXiv preprint arXiv:2210.13438}, 2022.

\bibitem[D{\'e}fossez et~al.(2024)D{\'e}fossez, Mazar{\'e}, Orsini, Royer, P{\'e}rez, J{\'e}gou, Grave, and Zeghidour]{moshi}
Alexandre D{\'e}fossez, Laurent Mazar{\'e}, Manu Orsini, Am{\'e}lie Royer, Patrick P{\'e}rez, Herv{\'e} J{\'e}gou, Edouard Grave, and Neil Zeghidour.
\newblock Moshi: a speech-text foundation model for real-time dialogue.
\newblock \emph{arXiv preprint arXiv:2410.00037}, 2024.

\bibitem[Dong et~al.(2024)Dong, Zhang, Zang, Cao, Wang, Ouyang, Zhang, Duan, Zhang, Li, et~al.]{internlm}
Xiaoyi Dong, Pan Zhang, Yuhang Zang, Yuhang Cao, Bin Wang, Linke Ouyang, Songyang Zhang, Haodong Duan, Wenwei Zhang, Yining Li, et~al.
\newblock Internlm-xcomposer2-4khd: A pioneering large vision-language model handling resolutions from 336 pixels to 4k hd.
\newblock \emph{arXiv preprint arXiv:2404.06512}, 2024.

\bibitem[Du et~al.(2024)Du, Chen, Zhang, Hu, Lu, Yang, Hu, Zheng, Gu, Ma, et~al.]{cosyvoice}
Zhihao Du, Qian Chen, Shiliang Zhang, Kai Hu, Heng Lu, Yexin Yang, Hangrui Hu, Siqi Zheng, Yue Gu, Ziyang Ma, et~al.
\newblock Cosyvoice: A scalable multilingual zero-shot text-to-speech synthesizer based on supervised semantic tokens.
\newblock \emph{arXiv preprint arXiv:2407.05407}, 2024.

\bibitem[Fang et~al.(2024)Fang, Guo, Zhou, Ma, Zhang, and Feng]{llama-omni}
Qingkai Fang, Shoutao Guo, Yan Zhou, Zhengrui Ma, Shaolei Zhang, and Yang Feng.
\newblock Llama-omni: Seamless speech interaction with large language models.
\newblock \emph{arXiv preprint arXiv:2409.06666}, 2024.

\bibitem[Fu et~al.(2024)Fu, Lin, Long, Shen, Zhao, Zhang, Wang, Yin, Ma, Zheng, et~al.]{vita}
Chaoyou Fu, Haojia Lin, Zuwei Long, Yunhang Shen, Meng Zhao, Yifan Zhang, Xiong Wang, Di~Yin, Long Ma, Xiawu Zheng, et~al.
\newblock Vita: Towards open-source interactive omni multimodal llm.
\newblock \emph{arXiv preprint arXiv:2408.05211}, 2024.

\bibitem[Google(2024)]{gemini}
Google.
\newblock https://deepmind.google/technologies/gemini/, 2024.

\bibitem[Li et~al.(2022)Li, Li, Xiong, and Hoi]{blip}
Junnan Li, Dongxu Li, Caiming Xiong, and Steven Hoi.
\newblock Blip: Bootstrapping language-image pre-training for unified vision-language understanding and generation.
\newblock In \emph{International conference on machine learning}, pages 12888--12900. PMLR, 2022.

\bibitem[Liu et~al.(2024)Liu, Li, Li, and Lee]{llava}
Haotian Liu, Chunyuan Li, Yuheng Li, and Yong~Jae Lee.
\newblock Improved baselines with visual instruction tuning.
\newblock In \emph{Proceedings of the IEEE/CVF Conference on Computer Vision and Pattern Recognition}, pages 26296--26306, 2024.

\bibitem[Ma et~al.(2024)Ma, Song, Du, Cong, Chen, Wang, Wang, and Chen]{LSLM}
Ziyang Ma, Yakun Song, Chenpeng Du, Jian Cong, Zhuo Chen, Yuping Wang, Yuxuan Wang, and Xie Chen.
\newblock Language model can listen while speaking.
\newblock \emph{arXiv preprint arXiv:2408.02622}, 2024.

\bibitem[meta(2024)]{llama3.1}
meta.
\newblock llama3.1, 2024.
\newblock URL \url{https://llama.meta.com/}.

\bibitem[Mihaiii(2024{\natexlab{a}})]{qa-assistant-2}
Mihaiii.
\newblock https://huggingface.co/datasets/mihaiii/qa-assistant-2, 2024{\natexlab{a}}.

\bibitem[Mihaiii(2024{\natexlab{b}})]{t-mul}
Mihaiii.
\newblock https://huggingface.co/datasets/mihaiii/triviamultichoice, 2024{\natexlab{b}}.

\bibitem[Mihaiii(2024{\natexlab{c}})]{t-sim}
Mihaiii.
\newblock https://huggingface.co/datasets/mihaiii/triviasinglechoice, 2024{\natexlab{c}}.

\bibitem[Nachmani et~al.(2023)Nachmani, Levkovitch, Hirsch, Salazar, Asawaroengchai, Mariooryad, Rivlin, Skerry-Ryan, and Ramanovich]{Spectron}
Eliya Nachmani, Alon Levkovitch, Roy Hirsch, Julian Salazar, Chulayuth Asawaroengchai, Soroosh Mariooryad, Ehud Rivlin, RJ~Skerry-Ryan, and Michelle~Tadmor Ramanovich.
\newblock Spoken question answering and speech continuation using spectrogram-powered llm.
\newblock \emph{arXiv preprint arXiv:2305.15255}, 2023.

\bibitem[Openai(2024{\natexlab{a}})]{gpt4o}
Openai.
\newblock https://openai.com/index/hello-gpt-4o/, 2024{\natexlab{a}}.

\bibitem[Openai(2024{\natexlab{b}})]{gpt4v}
Openai.
\newblock https://openai.com/index/gpt-4v-system-card/, 2024{\natexlab{b}}.

\bibitem[OpenAssistan(2024)]{open-assis}
OpenAssistan.
\newblock https://huggingface.co/datasets/openassistant/oasst1, 2024.

\bibitem[OpenOrca()]{openorca}
OpenOrca.
\newblock https://huggingface.co/datasets/open-orca/openorca/.

\bibitem[Pratap et~al.(2020)Pratap, Xu, Sriram, Synnaeve, and Collobert]{mls}
Vineel Pratap, Qiantong Xu, Anuroop Sriram, Gabriel Synnaeve, and Ronan Collobert.
\newblock Mls: A large-scale multilingual dataset for speech research.
\newblock \emph{arXiv preprint arXiv:2012.03411}, 2020.

\bibitem[Radford et~al.(2021)Radford, Kim, Hallacy, Ramesh, Goh, Agarwal, Sastry, Askell, Mishkin, Clark, et~al.]{clip}
Alec Radford, Jong~Wook Kim, Chris Hallacy, Aditya Ramesh, Gabriel Goh, Sandhini Agarwal, Girish Sastry, Amanda Askell, Pamela Mishkin, Jack Clark, et~al.
\newblock Learning transferable visual models from natural language supervision.
\newblock In \emph{International conference on machine learning}, pages 8748--8763. PMLR, 2021.

\bibitem[Radford et~al.(2023)Radford, Kim, Xu, Brockman, McLeavey, and Sutskever]{whisper}
Alec Radford, Jong~Wook Kim, Tao Xu, Greg Brockman, Christine McLeavey, and Ilya Sutskever.
\newblock Robust speech recognition via large-scale weak supervision.
\newblock In \emph{International conference on machine learning}, pages 28492--28518. PMLR, 2023.

\bibitem[Rubenstein et~al.(2023)Rubenstein, Asawaroengchai, Nguyen, Bapna, Borsos, Quitry, Chen, Badawy, Han, Kharitonov, et~al.]{audiopalm}
Paul~K Rubenstein, Chulayuth Asawaroengchai, Duc~Dung Nguyen, Ankur Bapna, Zal{\'a}n Borsos, F{\'e}lix de~Chaumont Quitry, Peter Chen, Dalia~El Badawy, Wei Han, Eugene Kharitonov, et~al.
\newblock Audiopalm: A large language model that can speak and listen.
\newblock \emph{arXiv preprint arXiv:2306.12925}, 2023.

\bibitem[sayan1101(2024)]{identity_finetune}
sayan1101.
\newblock https://huggingface.co/datasets/sayan1101/identity-finetune-data, 2024.

\bibitem[Siuzdak(2024)]{snac}
Hubert Siuzdak.
\newblock https://github.com/hubertsiuzdak/snac/, 2024.

\bibitem[Snyder et~al.(2015)Snyder, Chen, and Povey]{snyder2015musan}
David Snyder, Guoguo Chen, and Daniel Povey.
\newblock Musan: A music, speech, and noise corpus.
\newblock \emph{arXiv preprint arXiv:1510.08484}, 2015.

\bibitem[Sun et~al.(2024)Sun, Zhang, He, Li, Cheng, Liu, Yan, Shao, Tang, Zhang, et~al.]{moss}
Tianxiang Sun, Xiaotian Zhang, Zhengfu He, Peng Li, Qinyuan Cheng, Xiangyang Liu, Hang Yan, Yunfan Shao, Qiong Tang, Shiduo Zhang, et~al.
\newblock Moss: An open conversational large language model.
\newblock \emph{Machine Intelligence Research}, pages 1--18, 2024.

\bibitem[Touvron et~al.(2023)Touvron, Lavril, Izacard, Martinet, Lachaux, Lacroix, Rozi{\`e}re, Goyal, Hambro, Azhar, et~al.]{llama}
H~Touvron, T~Lavril, G~Izacard, X~Martinet, MA~Lachaux, T~Lacroix, B~Rozi{\`e}re, N~Goyal, E~Hambro, F~Azhar, et~al.
\newblock Open and efficient foundation language models.
\newblock \emph{Preprint at arXiv. https://doi. org/10.48550/arXiv}, 2302, 2023.

\bibitem[vicgalle(2024)]{Alpacagpt4}
vicgalle.
\newblock https://huggingface.co/datasets/vicgalle/alpaca-gpt4, 2024.

\bibitem[Wang et~al.(2023{\natexlab{a}})Wang, Chen, Wu, Zhang, Zhou, Liu, Chen, Liu, Wang, Li, et~al.]{valle}
Chengyi Wang, Sanyuan Chen, Yu~Wu, Ziqiang Zhang, Long Zhou, Shujie Liu, Zhuo Chen, Yanqing Liu, Huaming Wang, Jinyu Li, et~al.
\newblock Neural codec language models are zero-shot text to speech synthesizers.
\newblock \emph{arXiv preprint arXiv:2301.02111}, 2023{\natexlab{a}}.

\bibitem[Wang et~al.(2024)Wang, Bai, Tan, Wang, Fan, Bai, Chen, Liu, Wang, Ge, et~al.]{qwen2vl}
Peng Wang, Shuai Bai, Sinan Tan, Shijie Wang, Zhihao Fan, Jinze Bai, Keqin Chen, Xuejing Liu, Jialin Wang, Wenbin Ge, et~al.
\newblock Qwen2-vl: Enhancing vision-language model's perception of the world at any resolution.
\newblock \emph{arXiv preprint arXiv:2409.12191}, 2024.

\bibitem[Wang et~al.(2023{\natexlab{b}})Wang, Lv, Yu, Hong, Qi, Wang, Ji, Yang, Zhao, Song, et~al.]{cogvlm}
Weihan Wang, Qingsong Lv, Wenmeng Yu, Wenyi Hong, Ji~Qi, Yan Wang, Junhui Ji, Zhuoyi Yang, Lei Zhao, Xixuan Song, et~al.
\newblock Cogvlm: Visual expert for pretrained language models.
\newblock \emph{arXiv preprint arXiv:2311.03079}, 2023{\natexlab{b}}.

\bibitem[Xie and Wu(2024)]{mini-omni}
Zhifei Xie and Changqiao Wu.
\newblock Mini-omni: Language models can hear, talk while thinking in streaming.
\newblock \emph{arXiv preprint arXiv:2408.16725}, 2024.

\bibitem[Yang et~al.(2024)Yang, Yang, Hui, Zheng, Yu, Zhou, Li, Li, Liu, Huang, et~al.]{qwen2}
An~Yang, Baosong Yang, Binyuan Hui, Bo~Zheng, Bowen Yu, Chang Zhou, Chengpeng Li, Chengyuan Li, Dayiheng Liu, Fei Huang, et~al.
\newblock Qwen2 technical report.
\newblock \emph{arXiv preprint arXiv:2407.10671}, 2024.

\bibitem[Zen et~al.(2019)Zen, Dang, Clark, Zhang, Weiss, Jia, Chen, and Wu]{libritts}
Heiga Zen, Viet Dang, Rob Clark, Yu~Zhang, Ron~J Weiss, Ye~Jia, Zhifeng Chen, and Yonghui Wu.
\newblock Libritts: A corpus derived from librispeech for text-to-speech.
\newblock \emph{arXiv preprint arXiv:1904.02882}, 2019.

\bibitem[Zhan et~al.(2024)Zhan, Dai, Ye, Zhou, Zhang, Liu, Zhang, Yuan, Zhang, Li, et~al.]{anygpt}
Jun Zhan, Junqi Dai, Jiasheng Ye, Yunhua Zhou, Dong Zhang, Zhigeng Liu, Xin Zhang, Ruibin Yuan, Ge~Zhang, Linyang Li, et~al.
\newblock Anygpt: Unified multimodal llm with discrete sequence modeling.
\newblock \emph{arXiv preprint arXiv:2402.12226}, 2024.

\bibitem[Zhang et~al.(2023{\natexlab{a}})Zhang, Li, Zhang, Zhan, Wang, Zhou, and Qiu]{speechgpt}
Dong Zhang, Shimin Li, Xin Zhang, Jun Zhan, Pengyu Wang, Yaqian Zhou, and Xipeng Qiu.
\newblock Speechgpt: Empowering large language models with intrinsic cross-modal conversational abilities.
\newblock \emph{arXiv preprint arXiv:2305.11000}, 2023{\natexlab{a}}.

\bibitem[Zhang et~al.(2023{\natexlab{b}})Zhang, Zhang, Li, Zhou, and Qiu]{speechtokenizer}
Xin Zhang, Dong Zhang, Shimin Li, Yaqian Zhou, and Xipeng Qiu.
\newblock Speechtokenizer: Unified speech tokenizer for speech large language models.
\newblock \emph{arXiv preprint arXiv:2308.16692}, 2023{\natexlab{b}}.

\bibitem[Zhang et~al.(2023{\natexlab{c}})Zhang, Han, Qin, Wang, Bapna, Chen, Chen, Li, Axelrod, Wang, et~al.]{googleusm}
Yu~Zhang, Wei Han, James Qin, Yongqiang Wang, Ankur Bapna, Zhehuai Chen, Nanxin Chen, Bo~Li, Vera Axelrod, Gary Wang, et~al.
\newblock Google usm: Scaling automatic speech recognition beyond 100 languages.
\newblock \emph{arXiv preprint arXiv:2303.01037}, 2023{\natexlab{c}}.

\bibitem[Zhu et~al.(2023)Zhu, Chen, Shen, Li, and Elhoseiny]{minigpt4}
Deyao Zhu, Jun Chen, Xiaoqian Shen, Xiang Li, and Mohamed Elhoseiny.
\newblock Minigpt-4: Enhancing vision-language understanding with advanced large language models.
\newblock \emph{arXiv preprint arXiv:2304.10592}, 2023.

\end{thebibliography}
\end{document}